2021

# Modeling ADHD in Drosophila: Investigating the Effects of Glucose on Dopamine Production Demonstrated by Locomotion

MYUNG SUH CHOI

**Abstract**

Hyperactivity is one of the hallmarks of attention-deficit/hyperactivity disorder (ADHD), which is a common, highly heritable neuropsychiatric disorder. Aberrant dopamine signaling is a major theme in ADHD, and dopamine production is directly linked to the intensity and persistence of hyperactive conduct. The strength and persistence of hyperactivity responses in *Drosophila* to startle stimuli were measured in a study to determine the effects of sugar on dopamine development. The effects of sugar on dopamine production were modeled using *Drosophila melanogaster*. A total of four experimental groups, namely 1%, 3%, and 5% glucose, as well as a control group (without glucose) were taken for the diet of *Drosophila*, and these four different amounts of glucose were introduced to the growth medium where *Drosophila* was cultured. The movements of *Drosophila* in the four treatment groups were captured using a camera. This experiment was carried out five times, each time using a different batch of *Drosophila*. Each group's average velocity over time was also reported. The web adaptation of the *Drosophila* Activity Monitor (DAM) device was used to analyze the captured movies from the camera. The discrepancy between the test groups' resting velocities and the control group's resting velocities was statistically significant ($p<0.05$), while the $\Delta V$ values of the 5% glucose and 3% glucose groups were statistically distinct from the control group and one another ($p<0.05$) in terms of hyperactivity intensity but the difference in $\Delta V$ values between the 1% glucose and control groups, however, was not statistically significant ($p<0.05$). Furthermore, when it came to hyperactivity persistence, all four treatment classes were statistically different ($p<0.05$). Since the strength and persistence of hyperactive behavior are directly correlated to dopamine output, this study shows that higher glucose intake is associated with more hyperactivity, for both the intensity ($\Delta V$) and persistence (cool down time).

**Introduction**

ADHD (attention deficit hyperactivity disorder) is a common neuropsychiatric disorder marked by age-inappropriate, persistent hyperactivity and impulsivity, as well as difficulties focusing attention (Annon., 2000). The condition affects 5–6% of children worldwide, as well as 2.5 percent of adults (Franke *et al.*, 2012). ADHD is a neuropsychiatric disorder related to an attenuated and dysfunctional dopamine system. The disorder is characterized by impulsivity, hyperactivity, and short attention span. Humans with the disorder often take medication that increase levels of brain dopamine in order to reduce hyperactivity. ADHD is one of the most heritable neuropsychiatric conditions, with a heritability of 76 percent (Faraone *et al.*, 2005). The majority of studies have looked into genes that regulate dopamine homeostasis to see if they play a role in the disorder (Bralten *et al.*, 2013). Several meta-analyses found a correlation between ADHD and the gene encoding the dopamine transporter (Gizer *et al.*, 2009).

While studies have suggested that sugar is positively associated with a higher level of attention deficiency similar to ADHD symptoms, such studies were mostly retrospective, survey based observational studies, leaving little experimental evidence demonstrating an association between sugar consumption and dopamine production in the brain. Several studies have been conducted to show the genetic factors that cause ADHD symptoms in virtually all patients, but the cause remains unknown. To get around this, you'll need an efficient model that allows you to investigate specific usable data in a short period of time.

Though animal models are an excellent way to research the in *vivo* effects of altered gene function, there are currently only a few models for ADHD. Slc6a3 and Snap25 mutant mice have been observed the most (Wilson, 2000). In a zebrafish model of lphn3.1 downregulation, the locomotor phenotype is hyperactive/impulsive, with extreme reduction and misplacement of dopamine-positive neurons in the ventral diencephalon (Lange *et al.*, 2012). Lphn3 null mice have a hyperactive phenotype, which is followed by elevated dopamine and serotonin levels in the dorsal striatum (Wallis *et al.*, 2012). But there is no available model for ADHD to demonstrate the effect of sugar on ADHD yet.

The use of a genetic model in conjunction with effective methods and extensive services has the potential to make substantial progress in our understanding of the biological, cellular, and developmental basis of ADHD.

*Drosophila melanogaster* is a low-cost genetic model with a wide range of behaviors and tools that can be used to manipulate any gene of interest (Voet *et al.*, 2014). *Drosophila* is a valuable method for researching human brain diseases as a result of this. *Drosophila*'s suitability for modeling ADHD is backed up by research into the role of dopamine signaling in the fly's behavioral performance. A hyperactive mutant with a mutation in the homolog of the human ADHD-associated dopamine transporter gene has been identified (Kume *et al.*, 2005).

Lebestky *et al.* (2009) reports that closely paced, repetitive startle stimulus produces an extended period of hyperactivity in *Drosophila*. Studies have found that flies with an abnormally exaggerated hyperactivity response had a mutation in a dopamine receptor, and that flies with this dopamine-receptor mutation were hypersensitive to the air puffs, and took longer to calm down than "normal" flies without the mutation. Studies conclude that the intensity and persistence of the hyperactive behavior is directly correlated to dopamine production in the neural system in *Drosophila*. Voet *et al.* (2016) notes that the way the mutant flies respond to the air puffs is reminiscent of the way in which individuals with ADHD display hypersensitivity to environmental stimuli, strengthening the analogy between dopamine production in *Drosophila* and ADHD. Here in this study we use *Drosophila melanogaster* to model the effects of sugar on dopamine production, quantified by the intensity and persistence of hyperactivity response in *Drosophila* to startle stimuli.

**Materials and Methods**

This study used *Drosophila melanogaster* to model the effects of sugar on dopamine production, quantified by the intensity and persistence of hyperactivity response in *Drosophila* to startle stimuli.

Flies were raised on varying levels of added glucose in their diets. Using an air pump, tubes, and plastic connectors, a device that delivers a series of brief air puffs (startle stimuli) to horizontal vials containing *Drosophila* was built. Test groups then received the startle stimuli. The average velocity of each test group was analyzed via a web adaptation of the *Drosophila* Activity Monitor (DAM) system.

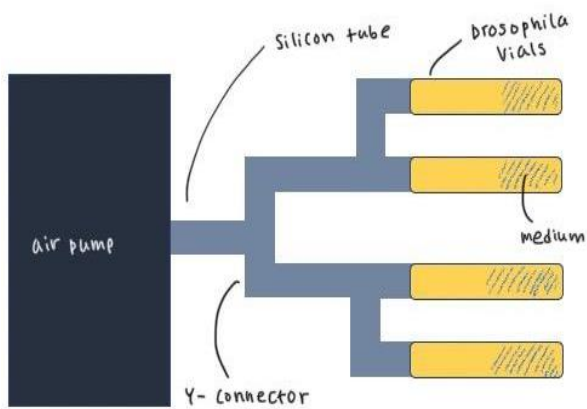

Figure: The stimulation apparatus

*Drosophila culture*

Three experimental groups were raised with varying levels of added glucose in their diets. Glucose was added to the growth medium by 1%, 3% and 5% by dry weight. The control group was raised on a regular growth medium, with no added glucose. Larvae were raised at room temperature and given seven days to fully mature.

*Data Collection*

*Drosophila* were temporarily anesthetized with $CO_2$ produced from an Alka-Seltzer tablet and a small amount of distilled water. 42 *Drosophila* from each test group were moved into the test vials.

A camera was set up so that all four *Drosophila* vials (1%, 3%, 5%, control) were against a plain white background and under a birds-eye view. 7, brief air puffs with a 5 second inter-puff interval were given to the 4 vials. Activity was recorded from one minute before the air-puff stimuli to 4 minutes after stimulus termination. This experiment was repeated 5 times, each with a new batch of *Drosophila*.

*Quantification*

The recorded movies were analyzed with a web adaptation of the *Drosophila* Activity Monitor (DAM) system. The average velocity for each group over time was recorded.

## Results

### *Rest velocity*

Even before the startle stimuli was administered, there was a difference in resting velocity among the groups. The average resting velocities, calculated from the first minute of observation prior to the startle stimuli, were 2.08 mm/s, 1.58 mm/s, 1.05 mm/s and 0.88 mm/s for the 5%, 3%, 1% and control group (0%), respectively. The difference between the test groups' resting velocities and the control group's resting velocity was statistically significant ($p<0.05$).

### *Hyperactivity: intensity*

Intensity of the Hyperactivity response was determined by calculating the difference between the peak velocity and the initial resting velocity. This value, $\Delta V$, was 8.47 mm/s, 7.74 mm/s, 6.29 mm/s and 6.24 mm/s for the 5%, 3%, 1% and control groups, respectively. The $\Delta V$ value of the 5% glucose and 3% glucose groups were statistically distinct from the control group and one another ($p<0.05$). The difference between the $\Delta V$ value of the 1% glucose group and the control group was not statistically significant ($p>0.05$).

### *Hyperactivity: persistence*

Persistence of the Hyperactivity response was determined by calculating the cool down time for each experimental group and control group. This value was 215 s, 190 s, 130 s, and 115 s for the 5%, 3%, 1% and control group (0%), respectively. All four values were statistically significant from one another ($p<0.05$).

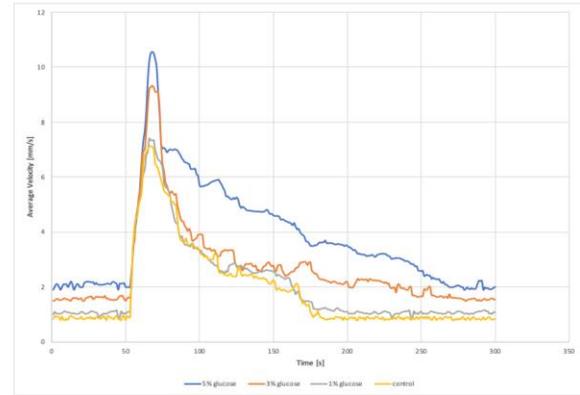

Figure 1. The average velocity (mm/s) for the three test groups (1%, 3%, and 5%) and the control group (0%) was calculated and recorded for 5 minutes. Each group demonstrated a similar pattern of activity, hitting a peak velocity and calming down to their initial rest velocity over a period of time.

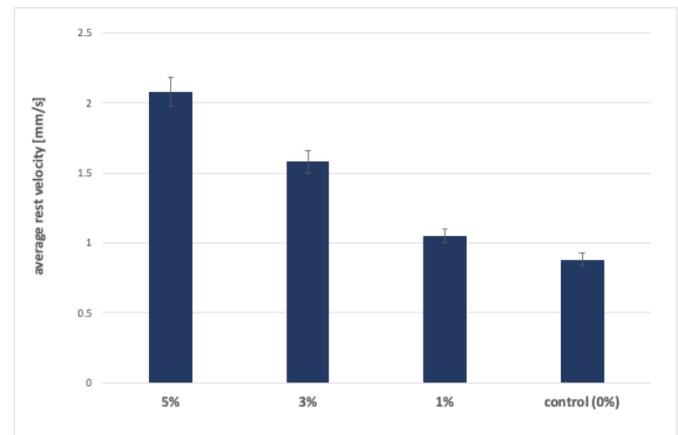

Figure 2. Average rest velocity was calculated for each group over the first minute of observation, prior to receiving the startle stimuli (air puffs).

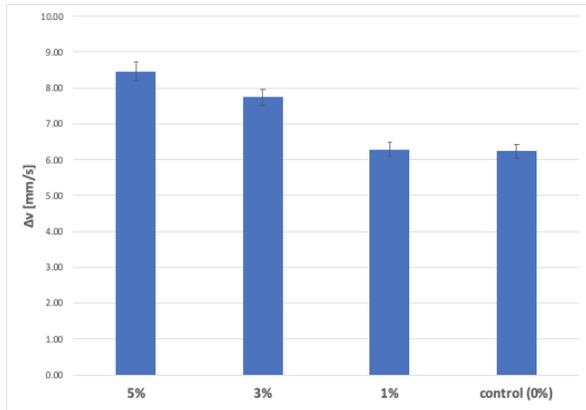

Figure 3. ΔV was calculated by subtracting the average rest velocity from the peak velocity.

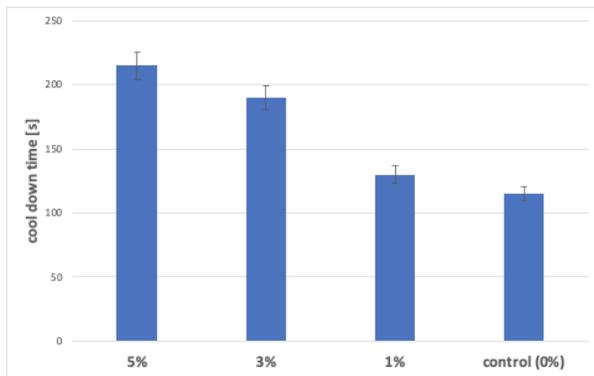

Figure 4. The cool down time was calculated by measuring the time it took from the test groups to hit their peak velocity to the first time they return to their average rest velocity (t at returned average rest velocity (mm/s) - t at peak velocity)

**Discussion**

While there has been extensive research done on the relationship between sugar consumption and ADHD, there is little experimental data that directly demonstrates a biological association between chronic sugar consumption and dopamine production. In this study, flies raised on a higher glucose diet demonstrated exaggerated hyperactivity compared to flies raised on a lower glucose diet. The 5% and 3% glucose-diet groups demonstrated hyperactivity with significantly higher intensity and persistence compared to the control group (0%). This study also shows that higher glucose consumption is associated with more hyperactivity, for both the intensity (ΔV) and persistence (cool down time) increased across the 0%, 1%, 3% and 5% glucose groups. Because the intensity and persistence of the hyperactive behavior is directly correlated to dopamine production in the neural system in *Drosophila*, conclusively, this study demonstrates that more glucose consumption is associated with less dopamine production in the neural system.

There was no statistically significant difference between the 1%-glucose diet group and the control group (0%) during the intensity assay. However, there was a statistically significant difference in the rest velocities (v) of the two groups, as well as in their cool down times. Studies suggest that early-onset ADHD is characterized by inattention and hyperactivity, while severe cases are diagnosed based hyperactive-impulsive symptoms. Paralleling such, while the 1% glucose group did not demonstrate statistically significant results regarding the intensity assay, which mirrors impulsive response symptoms of severe ADHD, the 1% glucose group did show significant initial restlessness (resting velocity) and later hyperactivity persistence (cool down time) compared to the control group (0%), mirroring the symptoms of early onset ADHD.

Contrastingly, the 3% and 5% groups produced significant results during the intensity assay, mirroring symptoms of more severe ADHD.

Lebansky *et al.* (2009) treated *Drosophila* with cocaine, a known dopamine agonist. The results from Lebansky's assay demonstrated that higher cocaine consumptions were positively correlated with a more exaggerated hyperactivity response in *Drosophila*, similar to how this study concluded that higher glucose consumption is positively correlated with a more exaggerated hyperactivity response in *Drosophila*. Such further supports the notion that chronic sugar consumption may directly alter dopamine levels within the neural system. Millichap *et al.* (2012) additionally reports that glucose directly affects physiological and pathological brain function, further supporting the hypothesis of this study. ADHD is a common neuropsychiatric disorder of major socioeconomic importance, impacting patients as young as infants to adults. Its etiology and neurobiology are poorly understood, and the relevance of high sugar diets and the biological consequences remain to be discovered. Studies have shown that *Drosophila* yields an ADHD-relevant, specific and readily recognizable locomotion phenotype that is indicative of dysregulated signaling in the dopaminergic circuit. *Drosophila* is a versatile and fast, cheap organism, and is a good model to dissect the mechanisms and pathways from gene to disease, in particular those associated with dopamine-related tracks. With hyperactivity as a start using *Drosophila* as a model, future studies exploring additional hallmarks of ADHD, such as cognitive capabilities, sensitization, and attention defects, through various mechanisms such as using an optomotor maze or other performance assays would be highly useful complements to this study.

**Conclusion**

In conclusion, flies like *Drosophila* raised on a higher glucose diet demonstrated exaggerated hyperactivity compared to flies raised on a lower glucose diet. This study also shows that higher glucose consumption is associated with more hyperactivity, for both the intensity ($\Delta V$) and persistence (cool down time) with increasing the glucose level. Because the intensity and persistence of the hyperactive behavior is directly correlated to dopamine production in the neural system in *Drosophila*. More glucose consumption is associated with less dopamine production in the neural system. This study also may propose that *Drosophila* may be a good model for dissecting mechanisms and pathways from gene to disease, especially those linked to dopamine-related genes, and drug sensitivity makes *Drosophila* models a good tool for finding new drugs leads because *Drosophila* is a flexible, fast, and low-cost organism for testing novel candidates identified in large-scale genetics studies in humans.

**Reference**

Alphen, B. V., & Swinderen, B. V. (2013). Drosophila strategies to study psychiatric disorders. Brain Research Bulletin, 92, 1-11. doi:10.1016/j.brainresbull.2011.09.007, read 10/30/20

Anjum, I., Jaffery, S. S., Fayyaz, M., Wajid, A., & Ans, A. H. (2018). Sugar Beverages and Dietary Sodas Impact on Brain Health: A Mini Literature Review. Cureus. doi:10.7759/cureus.2756 read 10/29/20

Annonymous. (2000). American Psychiatric Association Diagnostic and Statistical Manual of Mental Disorders. 4th edn American Psychiatric Association: Arlington.

Bralten, J., Franke, B., Waldman, I., Rommelse, N., Hartman, C., Asherson, P., Banaschewski, T., Ebstein, R.P., Gill, M., Miranda, A. and Oades, R.D. (2013). Candidate genetic pathways for attention-deficit/hyperactivity disorder (ADHD) show association to hyperactive/impulsive symptoms in children with ADHD. Journal of the American Academy of Child & Adolescent Psychiatry, 52(11), 1204-1212.

Dirks, H., Scherbaum, N., Kis, B., & Mette, C. (2017). ADHS im Erwachsenenalter und substanzbezogene Störungen – Prävalenz, Diagnostik und integrierte Behandlungskonzepte. Fortschritte Der Neurologie • Psychiatrie, 85(06), 336-344. doi:10.1055/s-0043-100763, read 10/29/20.

Dirks, H., Scherbaum, N., Kis, B., & Mette, C. (2017). Adult ADHD and Substance-Related Disorders, Prävalenz, Diagnostik und integrierte Behandlungskonzepte. Fortschritte Der Neurologie • Psychiatrie, 85(06), 336-344. doi:10.1055/s-0043-100763, read 10/29/20.

Faraone, S. V., Perlis, R. H., Doyle, A. E., Smoller, J. W., Goralnick, J. J., Holmgren, M. A., & Sklar, P. (2005). Molecular genetics of attention-deficit/hyperactivity disorder. Biological psychiatry, 57(11), 1313-1323.

Franke, B., Faraone, S.V., Asherson, P., Buitelaar, J., Bau, C.H.D., Ramos-Quiroga, J.A., Mick, E., Grevet, E.H., Johansson, S., Haavik, J. and Lesch, K.P. (2012). The genetics of attention deficit/hyperactivity disorder in adults, a review. Molecular psychiatry, 17(10), 960-987.

Gizer, I. R., Ficks, C., & Waldman, I. D. (2009). Candidate gene studies of ADHD: a meta-analytic review. Human genetics, 126(1), 51-90.

Klein, M., Singgih, E. L., Rens, A. V., Demontis, D., Børglum, A. D., Mota, N. R., . . . Franke, B. (2020). Contribution of Intellectual Disability–Related Genes to ADHD Risk and to Locomotor Activity in Drosophila. American Journal of Psychiatry, 177(6), 526-536. doi:10.1176/appi.ajp.2019.18050599, read 10/30/20.

Kume, K., Kume, S., Park, S. K., Hirsh, J., & Jackson, F. R. (2005). Dopamine is a regulator of arousal in the fruit fly. Journal of Neuroscience, 25(32), 7377-7384.

Lange, M., Norton, W., Coolen, M., Chaminade, M., Merker, S., Proft, F., Schmitt, A., Vernier, P., Lesch, K.P. and Bally-Cuif, L. (2012). The ADHD-susceptibility gene lphn3. 1 modulates dopaminergic neuron formation and locomotor activity during zebrafish development. Molecular psychiatry, 17(9), 946-954.

Lebestky, T., Chang, J. C., Dankert, H., Zelnik, L., Kim, Y., Han, K., . . . Anderson, D. J. (2009). Two Different Forms of Arousal in Drosophila Are Oppositely Regulated by the Dopamine D1 Receptor Ortholog DopR via Distinct Neural Circuits. Neuron, 64(4), 522-536. doi:10.1016/j.neuron.2009.09.031, read 10/30/20.

Millichap, J. G., & Yee, M. M. (2012). The Diet Factor in Attention-Deficit/Hyperactivity Disorder. Pediatrics, 129(2), 330-337. doi:10.1542/peds.2011-2199, read 10/29/20

van der Voet, Á., Harich, Á., Franke, Á., & Schenck, A. (2016). ADHD-associated dopamine transporter, latrophilin and neurofibromin share a dopamine-related locomotor signature in *Drosophila*. Molecular psychiatry, 21(4), 565-573.

van der Voet, M., Nijhof, B., Oortveld, M. A., & Schenck, A. (2014). *Drosophila* models of early onset cognitive disorders and their clinical applications. Neuroscience & Biobehavioral Reviews, 46, 326-342.

Voet, M. V., Harich, B., Franke, B., & Schenck, A. (2015). ADHD-associated dopamine transporter, latrophilin and neurofibromin share a dopamine-related locomotor signature in Drosophila. Molecular Psychiatry, 21(4), 565-573. doi:10.1038/mp.2015.55, read 11/1/20.

Wallis, D., Hill, D. S., Mendez, I. A., Abbott, L. C., Finnell, R. H., Wellman, P. J., & Setlow, B. (2012). Initial characterization of mice null for Lphn3, a gene implicated in ADHD and addiction. Brain research, 1463, 85-92.

Wilson, M. C. (2000). Coloboma mouse mutant as an animal model of hyperkinesis and attention deficit hyperactivity disorder.

Neuroscience & Biobehavioral Reviews, 24(1), 51-57.

Yildirim, K., Petri, J., Kottmeier, R., & Klämbt, C. (2018). Drosophila glia: Few cell types and many conserved functions. Glia, 67(1), 5-26. doi:10.1002/glia.23459, rea.